\documentclass[10pt,twocolumn,letterpaper]{article}

\usepackage{abcd}
\usepackage{times}
\usepackage{epsfig}
\usepackage{graphicx}
\usepackage{amsmath}
\usepackage{amssymb}
\usepackage{xcolor}
\usepackage{subcaption}
\usepackage{adjustbox}
\usepackage{lipsum}
\usepackage{authblk}

\usepackage{float}

\usepackage{multicol}
\usepackage{algorithm}
\usepackage[noend]{algpseudocode}

\usepackage[aboveskip=0pt]{caption}
\usepackage[pagebackref=true,breaklinks=true,letterpaper=true,colorlinks,bookmarks=false]{hyperref}

\abcdfinalcopy 
\def\abcdPaperID{****}

\ifabcdfinal\fi

\begin{document}

\title{Distortion Agnostic Deep Watermarking}

\author[1]{Xiyang Luo}
\author[2\thanks{Work done while an intern at Google.}]{Ruohan Zhan}
\author[1]{Huiwen Chang}
\author[1]{Feng Yang}
\author[1]{Peyman Milanfar}
\affil[1]{Google Research}
\affil[2]{Stanford University}
\date{}                     

\maketitle

\begin{abstract}
Watermarking is the process of embedding information into an image that can survive under distortions, while requiring the encoded image to have little or no perceptual difference from the original image. Recently, deep learning-based methods achieved impressive results in both visual quality and message payload under a wide variety of image distortions. However, these methods all require differentiable models for the image distortions at training time, and may generalize poorly to unknown distortions. This is undesirable since the types of distortions applied to watermarked images are usually unknown and non-differentiable. In this paper, we propose a new framework for distortion-agnostic watermarking, where the image distortion is not explicitly modeled during training. Instead, the robustness of our system comes from two sources: adversarial training and channel coding. Compared to training on a fixed set of distortions and noise levels, our method achieves comparable or better results on distortions available during training, and better performance on unknown distortions. 
\end{abstract}

\section{Introduction}
\label{sec:introduction}

Digital watermarking \cite{cox2002digital} is the task of embedding information into an image in a visually imperceptible fashion, where the message can be reliably extracted under image distortions. There are two key factors to measure the performance of a digital watermarking system, \emph{imperceptibility} and \emph{robustness}. Given an image and a message, a good watermarking system produces an encoded image that is nearly identical to the original image, while carrying a message payload that will survive under a variety of distortions such as cropping, blurrying, or JPEG compression. Traditional approaches found creative ways of hiding information in texture rich areas \cite{bender1996techniques} or the frequency domain \cite{kundur1998digital}.  More recently, convolutional neural networks (CNNs) have been used to provide an end-to-end solution to the watermarking problem. Zhu \textit{et al.} \cite{zhu2018hidden} proposed HiDDeN, a unified framework for digital watermarking and image steganography.

\begin{figure}
\vspace{-0 mm}
\begin{center}
\includegraphics*[width=.99\linewidth]{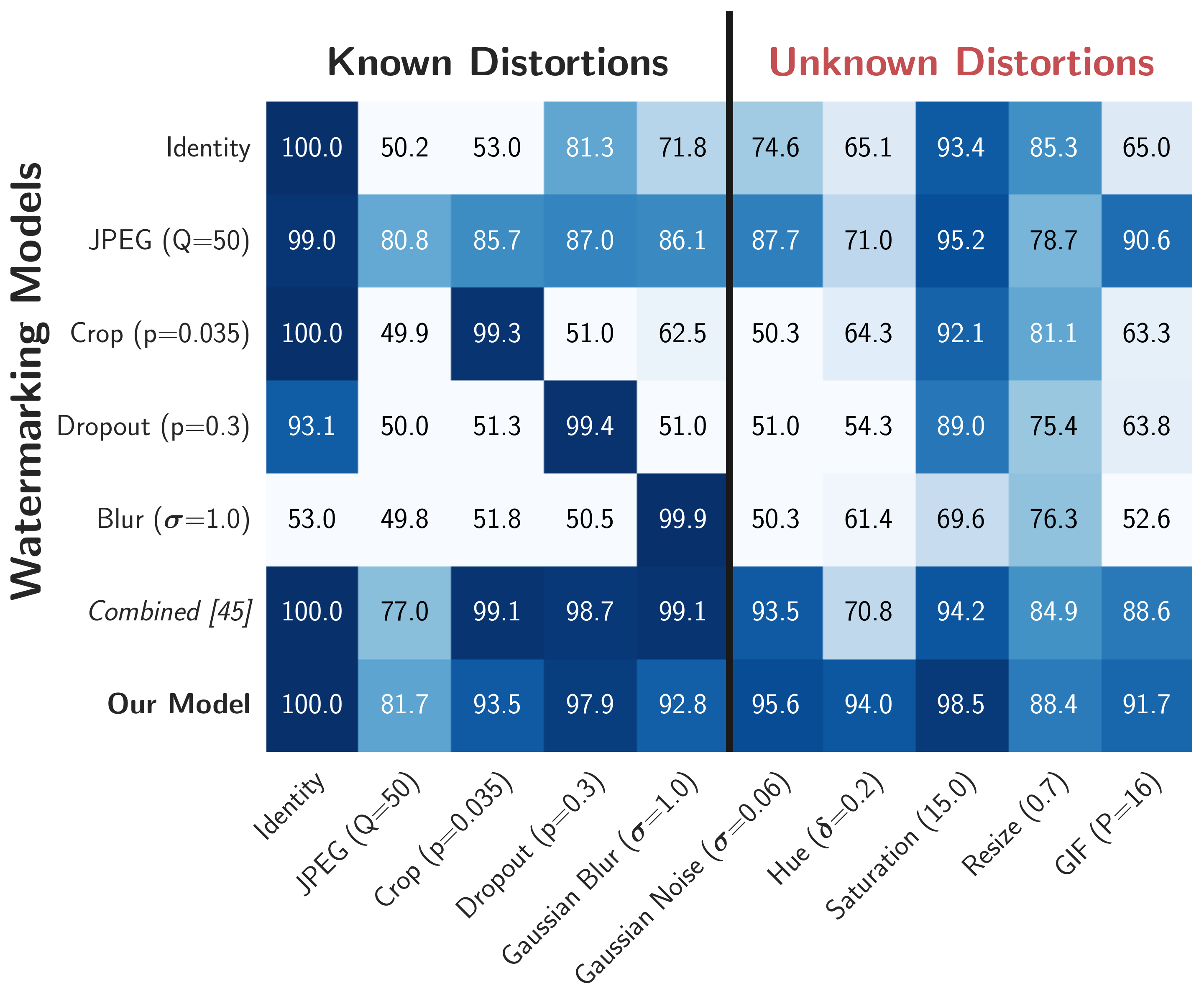}
\end{center}
{\caption{Bit accuracy of our model compared to models trained with explicit image distortions. Each column corresponds to a type of image distortion at test time, and each row corresponds to the image distortion used to train the watermarking model (with the exception of our model which requires no distortion model). The left half of the columns (separated by the black line) are \emph{known distortions}, \textit{i.e.}, distortions included in training for the HiDDeN combined model~\cite{zhu2018hidden}, and the right half of the columns \emph{unknown distortions}, \textit{i.e.}, a held-out set of commonly used distortions not used to train the HiDDeN combined model. See Section~\ref{subsec:comparison-hidden} for more details.} \label{fig:matrix_plot}}
\end{figure}


\begin{figure}
\begin{center}
    \begin{tabular}{c c c}
        \begin{subfigure}{.30\linewidth} \includegraphics[width=0.99\linewidth]{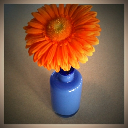} \end{subfigure}       
                 &  \begin{subfigure}{.30\linewidth} \includegraphics[width=0.99\linewidth]{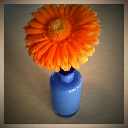} \end{subfigure}     
                 &  \begin{subfigure}{.30\linewidth} \includegraphics[width=0.99\linewidth]{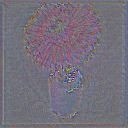} \end{subfigure}   \\
                 Original & Encoded & Difference 
    \end{tabular}
\end{center}
\caption{Example of original image, encoded image and difference between the two images from our model. }
\vspace{-5 mm} \label{fig:front-page-samples}
\end{figure}

\begin{figure*}
  \includegraphics[width=\textwidth]{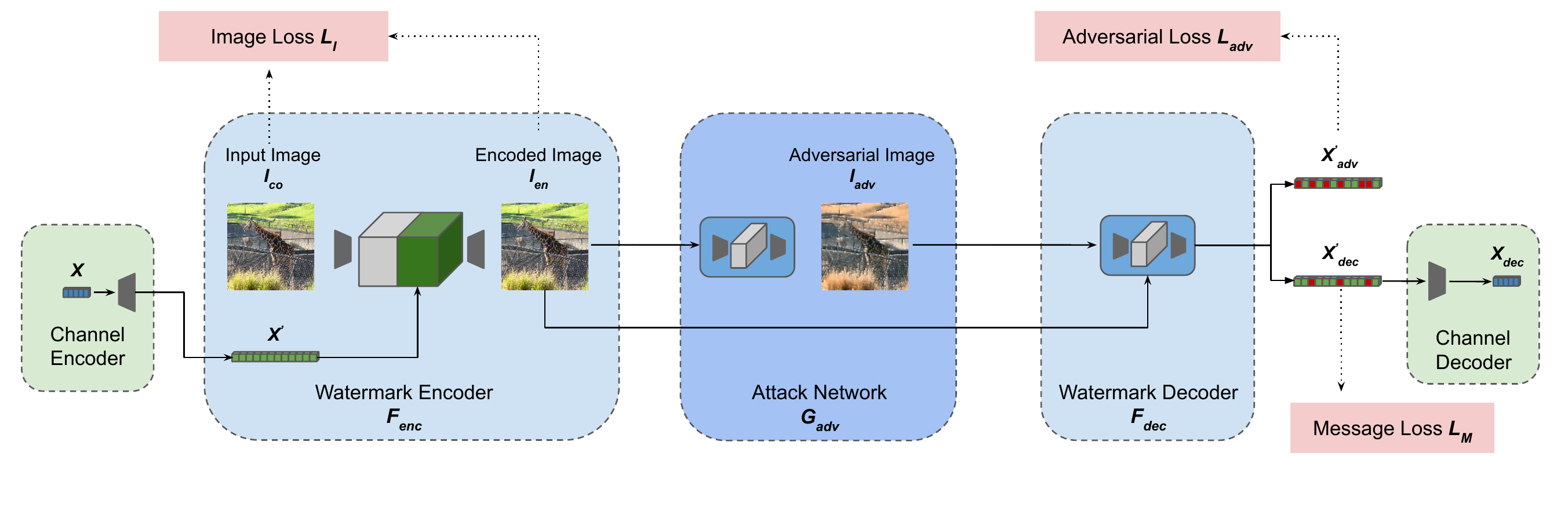}
  \caption{Overview of proposed architecture. The input message $X$ is first fed through the channel encoder to produce a redundant message $X'$, which is then combined with the input image $I_{co}$ to generate the encoded image $I_{en}$ by the watermark encoder $F_{enc}$. The decoder $F_{dec}$ produces a decoded message $X'_{dec}$, where it is further processed by the channel decoder to produce the final message $X_{dec}$. The attack network generates adversarial examples $I_{adv}$, which are fed to the image decoder to obtain $X'_{adv}$.  $F_{enc}$, $F_{dec}$, are trained on a combination of the image loss $L_I$ which includes both proximity to the cover image $I_{co}$ and perceptual quality as in Equation~\ref{eq:adv1}, the message loss $L_M$ as in Equation~\ref{eq:adv2}, and the message loss on the decoded adversarial message $X'_{adv}$ as in Equation~\ref{eq:adv4}. The attack network $G_{adv}$ is trained to minimize the adversarial loss $L_{adv}$ as in Equation~\ref{eq:adv3}. The training updates $G_{adv}$ and the $F_{enc}$, $F_{dec}$ in an alternating fashion. }
  \label{fig:overall_architecture}
\end{figure*}

 Most CNN based watermarking methods explicitly model the image distortions during training. However, training on a specific image distortion can easily lead to overfitting, and generalize poorly across other types of distortions~\cite{zhu2018hidden}. This is undesirable since a practical watermarking system should be robust towards a wide variety of image distortions, not only the ones included in training. This can be mitigated by including a combination of distortions during training~\cite{zhu2018hidden}, but it requires carefully tuning the type and magnitude of the distortions in order to reach a good performance. Moreover, the problem of poor generalization persists if the distortions at test time are far from training.

To this end, we propose a framework for adding robustness to a watermarking system \emph{without} any prior knowledge on the type of image distortions during training. We achieve this by applying differentiable adversarial training with CNN generated perturbations, and using channel coding to inject redundancy in the encoded message. To our knowledge, this is the first paper explore distortion agnostic methods for deep watermarking. Empirically, our model achieves comparable performance on distortions known during training, and generalization better to unknown distortions.

Our main contributions are the following. 
\begin{itemize}
    \item We apply adversarial training to improve model robustness in a distortion agnostic fashion. In particular, our CNN generated adversarial examples implicitly incorporates a rich collection of image distortions that co-adapt with training.
    \item We propose augmenting the watermarking system with channel coding, adding an additional layer of robustness through channel redundancy. 
    \item We combine the two ideas above and achieve comparable results to models trained with explicit distortion, and better generalization to unknown distortions.    
\end{itemize}

\section{Related Work}
\label{sec:related}
There are three main areas of research relevant to this work: watermarking, adversarial training, and channel coding. We give a brief review for each topic in the subsections below.  

\begin{figure*}
\begin{center}
\vspace{-0 mm}
    \begin{tabular}{c c c c c c}
        $I_{en}$  &  \begin{subfigure}{.14\linewidth} \includegraphics[width=0.99\linewidth]{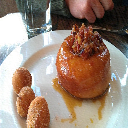} \end{subfigure}       
                 &  \begin{subfigure}{.14\linewidth} \includegraphics[width=0.99\linewidth]{./figures/adv_original_9.png} \end{subfigure}     
                 &  \begin{subfigure}{.14\linewidth} \includegraphics[width=0.99\linewidth]{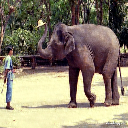} \end{subfigure}
                 &  \begin{subfigure}{.14\linewidth} \includegraphics[width=0.99\linewidth]{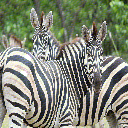} \end{subfigure}
                 &  \begin{subfigure}{.14\linewidth} \includegraphics[width=0.99\linewidth]{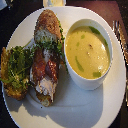} \end{subfigure}          \vspace{10pt}       \\    
        $I_{adv}$   &  \begin{subfigure}{.14\linewidth} \includegraphics[width=0.99\linewidth]{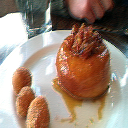} \end{subfigure}       
                 &  \begin{subfigure}{.14\linewidth} \includegraphics[width=0.99\linewidth]{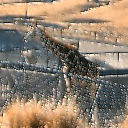} \end{subfigure}     
                 &  \begin{subfigure}{.14\linewidth} \includegraphics[width=0.99\linewidth]{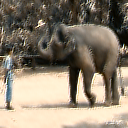} \end{subfigure}
                 &  \begin{subfigure}{.14\linewidth} \includegraphics[width=0.99\linewidth]{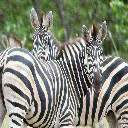} \end{subfigure}
                 &  \begin{subfigure}{.14\linewidth} \includegraphics[width=0.99\linewidth]{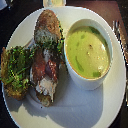} \end{subfigure}          \vspace{10pt}       \\   
        & (a) blur & (b) color and noise & (c) blur and color & (d) pattern & (e) color
    \end{tabular}
\end{center}
\caption{Visualization of adversarial examples generated by the attack network $G_{adv}$. \textbf{Top:} Encoded image $I_{en}$. \textbf{Bottom:} Adversarial examples $I_{adv}$ generated from the attack network $G_{adv}$. We observe a diverse set of image manipulations generated from the attack network, consisting of a combination of blur, color change, and other types of distortions. } \label{fig:adv_examples}
\end{figure*}

\subsection{Watermarking}
Digital watermarking \cite{cox2002digital, katzenbeisser2000digital,bender1996techniques,jiansheng2009digital,o1997rotation,singh2013survey,tanaka1990embedding,hamidi2015blind} have been an active research area with many important applications such as content copyright protection. More recently, deep learning based approaches have been applied to train an end-to-end watermarking system \cite{zhu2018hidden,mun2017robust,ahmadi2018redmark,liu2019novel, zhang2019steganogan} with impressive results. HiDDeN~\cite{zhu2018hidden} was one of the first deep learning solutions for image watermarking. RedMark~\cite{ahmadi2018redmark} introduced residual connections with a strength factor for embedding binary images in the transform domain. Deep watermarking has since been generalized to both video~\cite{weng2019high,zhang2019robust} and audio~\cite{tegendal2019watermarking}. Modeling more complex and realistic image distortions also broadened the scope in terms of application \cite{wengrowski2019light, tancik2019stegastamp}.

There are several works that applied attacks to the encoded image when training the watermarking system. Mun \textit{et al.}~\cite{mun2017robust} iteratively simulated attacks to the watermarking system. RedMark~\cite{ahmadi2018redmark} introduced an attack layer which consists of random combinations of a fixed set of distortions. However, these attacks are not adversarial since they do not adapt with the watermarking model during training. Recently, ROMark~\cite{wen2019romark} applied a simple form of adversarial training where the distortion type and distortion strength are adaptively selected to minimize the decoding accuracy.

One key distinction of our method from the above is that we do \emph{not} generate our attacks from a \emph{fixed} pool of common distortions. Instead, the adversarial examples are generated from a trained CNN. This also has the benefit that the watermarking training is end-to-end differentiable, which is not true for ROMark~\cite{wen2019romark}.

\subsection{Adversarial Training}
Deep neural networks are susceptible to certain tiny perturbations in the input space. Since the discovery of adversarial examples by Szegedy \textit{et al.} \cite{szegedy2013intriguing}, a variety of methods have been proposed for both adversarial attack~\cite{athalye2018obfuscated,kurakin2016adversarial,papernot2017practical}, and adversarial defense \cite{goodfellow2014explaining,hosseini2017blocking,shafahi2019adversarial,xie2017mitigating}. One of the earliest and most effective defense mechanism against adversarial attacks is adversarial training \cite{goodfellow2014explaining}, but is computationally expensive on large datasets. Many attempts have since been made to reduce the cost of adversarial training, \textit{e.g.}, using approximations to the optimization step~\cite{shafahi2019adversarial}, or using generative models in place of iterative optimization~\cite{baluja2017adversarial,lee2017generative}.

\subsection{Channel coding}
Channel coding is a mechanism for detecting and correcting errors during signal transmission \cite{bossert1999channel}. Shannon's capacity theorem~\cite{shannon1948mathematical} gives the theoretical limit to transfer data through a noisy channel, and channel coding is designed to approach this limit.
In implementation, various classical methods such as the Reed-Solomon (RS) codes~\cite{wicker1999reed}, low-density parity-check (LDPC) codes~\cite{ryan2004introduction}, turbo codes~\cite{sklar1997primer}, and polar codes~\cite{trifonov2012efficient}, have been widely applied in the field of telecommunication.
More recently, learning based solutions have gained attention in this field as well~\cite{aoudia2019model,choi2018necst,fritschek2019deep}. 

\section{Proposed Method}
\label{sec:method}

\subsection{Motivation}
In designing a general purpose watermarking model, the distortions at test time could be any image manipulation that still preserves some image content. A typical solution would involve identifying a set of representative distortions, and applying a carefully tuned combination of distortions during training.

Motivated by the recent success of using CNNs to perform various image manipulation tasks, \textit{e.g.}, style transfer~\cite{gatys2016image}, HDRNet~\cite{li2019hdrnet}, we propose automating the distortion tuning process by \emph{training} a CNN to generate distortions that exploits the weakest link in the current watermarking model. Figure~\ref{fig:adv_examples} shows some samples of distorted images generated by our attack CNN, which contain a rich and complex combination of distortions. 

The use of channel coding is motivated by the idea of injecting extra redundancy to the system. Shannon's capacity theorem tells us that redundancy is necessary in order to achieve robustness. In the HiDDeN architecture, spatially repeating the input message is an example of adding redundancy. Channel coding simply provides another alternative on top of the current methods.

\subsection{Method Overview}
Figure~\ref{fig:overall_architecture} gives an overview of our overall architecture. Our method adds two key components on top of the watermarking encoder / decoder networks $F_{enc}$ and $F_{dec}$ in ~\cite{zhu2018hidden}:
\begin{itemize}
\item We replace the distored image with $I_{adv}$, where $I_{adv}$ is an adversarial example generated from a convolutional neural network trained to \emph{maximize} the message loss.  

\item We replace the input message with a longer binary message $X'$ generated from channel coding. 
\end{itemize}


\subsection{Adversarial Training}
Adversarial training generates distortions that co-adapt with the training of our watermarking model, actively strengthening the weakest point of the current model. Adversarial training was first introduced by Goodfellow \textit{et al.}~\cite{goodfellow2014explaining} as a method to defend against adversarial attacks. In our context, adversarial training equates to minimizing the message loss given the worst-case distortion in an $\epsilon$-ball. This is expressed as the following min-max problem,
\begin{equation}
\label{eq:adv1}
    \min_{\Theta_{enc}, \Theta_{dec}} \max_{\|\delta\| \leq \epsilon }\{ L_M(F_{dec}(F_{enc}(I_{co}; \Theta_{enc}) + \delta; \Theta_{dec}), X) \},
\end{equation}
where $\Theta_{enc}, \Theta_{dec}$ are the model parameters for watermarking encoder/decoder networks $F_{enc}$, $F_{dec}$, and $X$ is the input message. Here we consider the $L_2$ norm $\|\cdot\|_2$ to constrain $\delta$, the perturbation to the encoded image $I_{en} = F_{enc}(I_{co}; \Theta_{enc})$. But more semantically meaningful measures such as $L_2$ distance on VGG~\cite{simonyan2014very} activations could also be used.

A direct optimization of Equation~\ref{eq:adv1} is both computationally expensive and overly restrictive for the watermarking model. Instead, we relax Equation~\ref{eq:adv1} by restricting the set of distortions $\delta$ to be generated from some class of convolutional neural network $G_{adv}(I; \Theta_{adv})$.
\begin{equation}
\label{eq:adv2}
    \min_{\{\Theta_{enc, dec}\}} \max_{\{\|G_{adv}(I_{en})-I_{en}\| \leq \epsilon \}}\{ L_M(F_{dec}(G_{adv}(I_{en})); X) \}.
\end{equation}
Using CNN generated adversarial examples have the benefit of retaining the ability to generate a diverse set of image distortions, as shown in Figure~\ref{fig:adv_examples}. An alternative is to generate the adversarial samples via the Fast Gradient Sign Method (FGSM) as in \cite{goodfellow2014explaining}. But we found this yielded less diverse examples compared to CNN generated examples, and resulted in poorer overall robustness against distortion.


To train the attack network $G_{adv}$, we minimize the following adversarial training loss:
\begin{equation}
\label{eq:adv3}
    L_{adv}=  \alpha^{adv}_1 \|I_{adv} - I_{en}\|^2 - \alpha^{adv}_2  L_M(F_{dec}(I_{adv}); X),
\end{equation}
where $I_{adv}=G_{adv}(I_{en})$ is the adversarial example, $L_M$ is the message loss which we set as the $L_2$ loss in this paper, and $\alpha^{adv}_1 $, $\alpha^{adv}_2 $ are the scalar weights. $\alpha^{adv}_1 $ controls the strength of the distortion generated by the attack network $G_{adv}$, while $\alpha^{adv}_2$ controls the strength of the message loss for $G_{adv}$.  

For the network $G_{adv}$, we use a two-layer CNN,
\begin{equation}
\label{eq:adv4}
   G_{adv}(I) = \text{Conv}_{3}\circ \text{Leaky ReLU} \circ \text{Conv}_{16} (I).
\end{equation}
In general, we find that finding the right balance of attack strength, controlled by the complexity of $G_{adv}$ and the ratio between $\alpha^{adv}_1$ and $\alpha^{adv}_2$, is important for training. An overly strong attack results in slow training and a failure of the watermarking network to adapt to the adversarial examples, while an overly simple attack results in less robustness of the trained model. A detailed analysis can be found in Section~\ref{subsec:detailed-analysis}.

\subsection{Channel coding}
\begin{figure}[!ht]
\vspace{-0 mm}
\begin{center}
\includegraphics*[width=\linewidth]{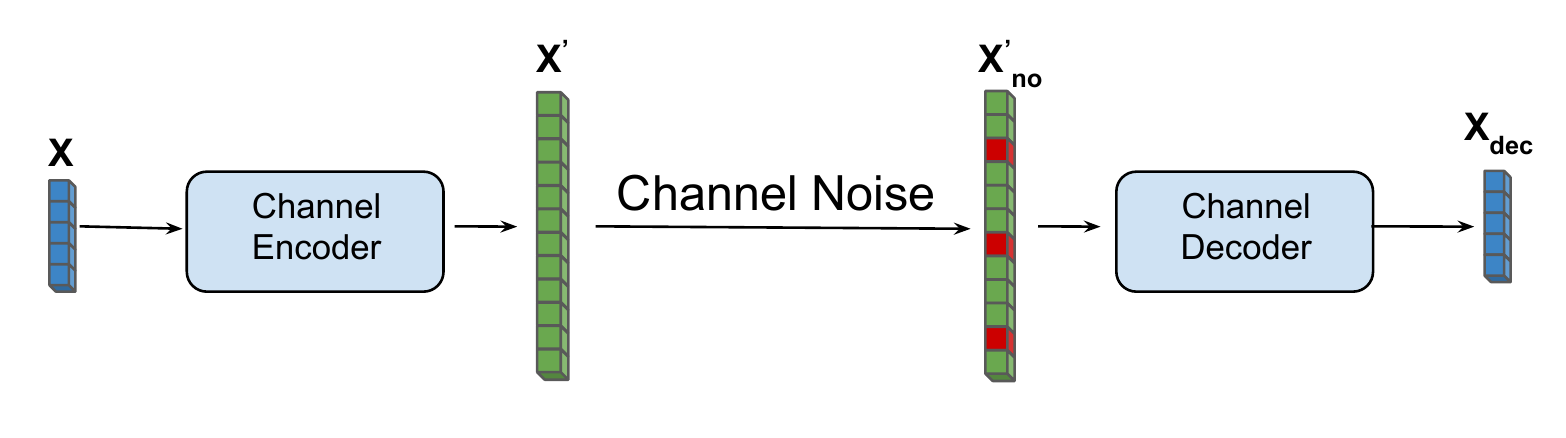}
\end{center}
\vspace{-2 mm}
{\caption{Illustration of channel coding. Given an input message $X$, the channel encoder produces a redundant message $X'$ of longer length. The redundant message $X'$ is transmitted through a noisy channel and received by the decoder as $X'_{no}$. Finally the decoder recovers the input $X$ from the corrupted message $X'_{no}$.} \label{fig:channel_coding}}
\vspace{-2 mm}
\end{figure}
Channel coding provides an additional layer of robustness through injecting redundancy to the system. Given a binary message $X\in\{0, 1\}^D$ of length $D$, a channel encoder produces a redundant message $X'\in\{0, 1\}^N$ of length $N>D$, which can be used to recover $X$ through the channel decoder given reasonable amounts of channel distortion to $X'$, as shown in Figure~\ref{fig:channel_coding}.

In this paper, we generate a channel code $X'$ from the input message $X$, before passing $X'$ to the watermarking encoder as shown in Figure~\ref{fig:overall_architecture}.   The channel distortions in this context are the errors from the watermarking model, between  $X'$ and $X'_{dec}$. Given that we do not explicitly model the image distortions, it is impossible to know the true channel distortion model. Instead, we use a binary symmetric channel (BSC) to approximate the channel distortion. BSC is a standard channel model which assumes each bit is independently and randomly flipped with probability $p$. Even though this assumption is not strictly satisfied in our case, we find using BSC works well in this application.

\begin{figure}[!ht]
\vspace{-0 mm}
\begin{center}
\includegraphics*[width=\linewidth]{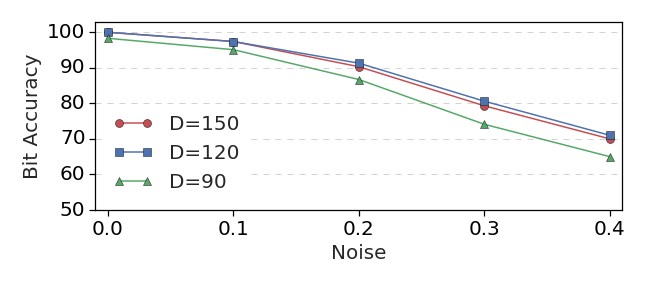}
\end{center}
\vspace{-2 mm}
{\caption{Channel noise strength versus decoder bit accuracy for various redundant message lengths. The input message length is fixed at $D=30$, where the redundant message length $N$ is varied from $90$ to $150$. All models are trained on random binary input with BSC noise. The training noise level is uniformly sampled from $[0, 0.3]$, and $[0, 0.4]$ at test time. } \label{fig:channel_bit_acc}}
\vspace{-2 mm}
\end{figure}

Conceptually, any standard error correcting code such as low-density parity-check (LDPC) codes~\cite{ryan2004introduction} can be used to generate $X'$. However, traditional codes such as LDPC require the decoder to have an estimate of the channel noise strength, which is impractical in our application since the noise strength can vary greatly from image to image. Therefore, we use NECST~\cite{choi2018necst}, a learning based solution for joint source and channel coding to cover a broad range of channel distortion strengths. We use BSC for training the channel model, where the input message $X$ is randomly sampled, and the channel noise strength is chosen from the interval $[0, \text{maxstrength}]$ uniformly at random. Figure~\ref{fig:channel_bit_acc} shows the bit accuracy of the NECST model on a range of BSC channel noise.

We emphasize here that the channel coding model is \emph{not} jointly trained with the rest of the watermarking model. This decoupling prevents the channel models from co-adapting with the image models during training, which results in overfitting and less robustness across a wide spectrum of image distortions.

\subsection{Watermarking Training and Losses}
We give a detailed description of the algorithms for training the watermarking models. We first define the training losses, using the same notations as in Figure~\ref{fig:overall_architecture}.

\textbf{Image loss}
\begin{equation}
\label{eq:adv-image-loss}
    L_I = \alpha^I_1 \|I_{co} - I_{en}\|^2 + \alpha^I_2 L_G(I_{en})
\end{equation}

\textbf{Message loss}
\begin{equation}
\label{eq:adv-message-loss}
    L_M = \alpha^M\|X'_{dec} - X'\|^2
\end{equation}

\textbf{Attack network training loss}
\begin{equation}
\label{eq:adv-attack-loss}
    L_{adv} = \alpha^{adv}_1 \|I_{adv} - I_{enc}\|^2 - \alpha^{adv}_2 \|X'_{adv} - X'\|^2
\end{equation}

\textbf{Watermarking training loss}
\begin{equation}
\label{eq:adv-watermark-loss}
    L_{W} = L_I + L_M + \alpha^{adv}_W \|X'_{adv} - X'\|^2
\end{equation}

The image loss in Equation~\ref{eq:adv-image-loss} consists of an $L_2$ loss, and a GAN loss $L_G$ with spectral normalization~\cite{miyato2018spectral} to control the perceptual quality of the encoded image. This is similar to the \emph{adversarial loss} defined in  the HiDDeN network \cite{zhu2018hidden}. For the message loss $L_M$, we use the $L_2$ loss between the decoded message and input. Equation~\ref{eq:adv-attack-loss} defines the loss used to train the attack network $G_{adv}$. Finally, Equation~\ref{eq:adv-watermark-loss} defines the overall loss for training $F_{enc}$ and $F_{dec}$. The various $\alpha$s are the weights for each loss. Training alternates between updating the attack network $G_{adv}$ and the watermarking networks $ F_{enc}, F_{dec}$, detailed in Algorithm~\ref{alg:watermark-train}. 


\begin{algorithm}
\caption{Watermarking Training}\label{alg:watermark-train}
\begin{algorithmic}[1]
\Procedure{Watermarking Train}{}\newline
\textbf{Input:} $I_{co}$, $X' \sim \textbf{Unif}(\{0, 1\}^N)$.\newline
\textbf{Output:} Trained networks $G_{adv}, F_{enc}, F_{dec}$.\newline
\textbf{Training Variables: } $\Theta_{adv}, \Theta_{enc}, \Theta_{dec}$.
\While{Step $<$ $max\_steps$}
  \State{Compute $I_{en} = F_{enc}(I_{co}, X)$}
  \For{i = 1 to $num\_iter$}
    \State{Compute $I_{adv} = G_{adv}(I_{en})$}  
    \State{Update $\Theta_{adv}= \Theta_{adv} + lr\times\text{Adam}(L_{adv})$}
  \EndFor
  \State{Update $\Theta_{dec}= \Theta_{dec} + lr\times\text{Adam}(L_{W})$}
  \State{Update $\Theta_{enc}= \Theta_{enc} + lr\times\text{Adam}(L_{W})$}
\EndWhile
\EndProcedure
\end{algorithmic}
\end{algorithm}



\section{Experiments}
\label{sec:experiments}
For comparison, we train two versions of HiDDeN~\cite{zhu2018hidden} as the baseline, one without image distortion which we name the \emph{identity} model, and another trained on a combination of standard image distortions which we name the \emph{combined} model. We note here that our methodology is agnostic to the specific architecture of the watermarking networks. We use the original HiDDeN architecture throughout the experiments since it is a well studied model and a commonly used benchmark, but other architectures such as RedMark~\cite{ahmadi2018redmark} could also be used as well.

We compare the bit accuracy on distortions seen during training and those that have not, and also report the peak signal-to-noise ratio (PSNR) of the encoded images.  All models are trained and evaluated on the MS COCO dataset~\cite{lin2014microsoft} resized to $128 \times 128$, where a random selection of 3000 images are used for evaluation. Unless otherwise stated, we use $D = 30$ for the encoded message size, and $N=120$ for the redundant message size. For the watermarking networks, we use the same architecture as used in HiDDeN, with the exception that the embedded message size is $120$ instead of $30$ due to the increased message length from channel coding. Detailed training parameters can be found in the supplementary materials.

\begin{figure*}
\vspace{-0 mm}
\begin{subfigure}{.49\linewidth}
    \includegraphics[width=0.99\linewidth]{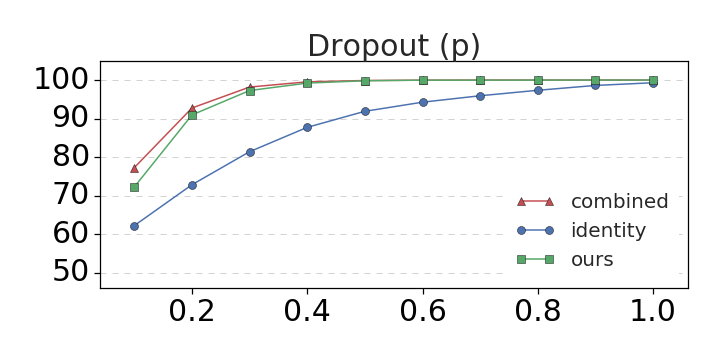}
\end{subfigure}
\begin{subfigure}{.49\linewidth}
    \includegraphics[width=0.99\linewidth]{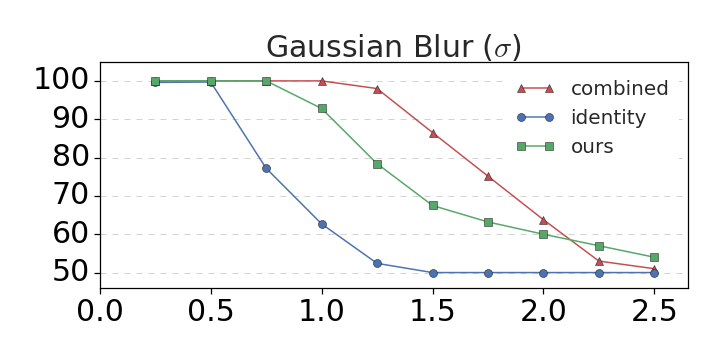}
\end{subfigure} \vspace{-10pt} \\ 
\begin{subfigure}{.49\linewidth}
    \includegraphics[width=0.99\linewidth]{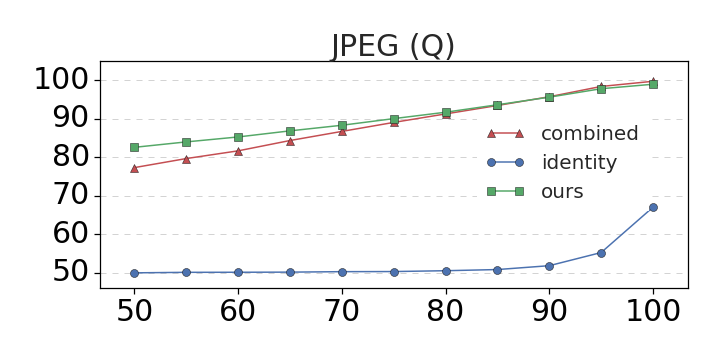}
\end{subfigure}
\begin{subfigure}{.49\linewidth}
    \includegraphics[width=0.99\linewidth]{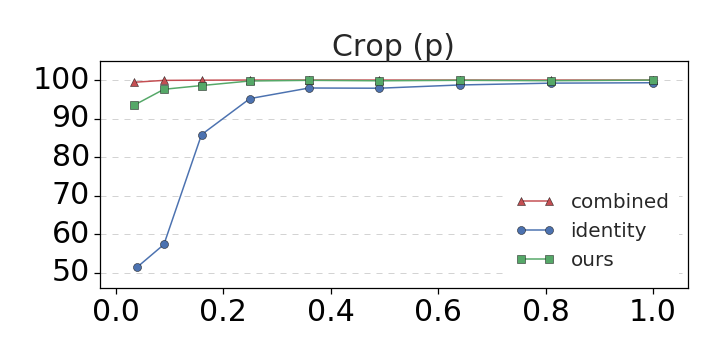}
\end{subfigure}
\caption{Comparison of our model with HiDDeN identity model and combined noise model for different types of image distortions.} \label{fig:noise_results}
\end{figure*}

\begin{table}[!ht]
    \begin{center}
    \begin{tabular}{c c c c }
    \hline
      Model&   Identity               & Combined     & Ours \\ \hline
        RGB- PSNR&       40.3       &  32.3            &  33.7                  \\     
        Y- PSNR&         47.5       &  34.2            &  35.7                   \\     
        U- PSNR&         44.5       &  39.7            &  40.7                     \\      
        V- PSNR&         43.1       &  39.5            &  40.1                     \\   \hline      
    \end{tabular}
    \end{center}
    \caption{Comparison of encoded image quality. The PSNR values in both RGB and YUV are reported for our model, as well as the HiDDeN identity and combined distortion model.}
    \label{tab:PSNR}
\end{table}

\subsection{Comparison with HiDDeN}
\label{subsec:comparison-hidden}

We compare our method with both the HiDDeN identity and combined models. For the combined model, we use JPEG ($q=50$), dropout ($q=0.3$), crop ($p=0.3$), Gaussian blur ($\sigma=1.0$), where the JPEG distortion is approximated by the differentiable JPEG function~\cite{shin2017jpeg}.  We also compare with \emph{specialized} models trained only on a single type of distortion with the noise levels in Figure~\ref{fig:matrix_plot}. For a fair comparison, we adjust $\alpha^I_1$ to obtain a slightly higher PSNR compared to the combined model, as shown in Table~\ref{tab:PSNR}.

Figure~\ref{fig:matrix_plot} shows the bit accuracy of our model and those trained with explicit image distortion. Each row corresponds to a different watermarking model, and each column a specific type of distortion applied at evaluation time. The top five rows (specialized models) clearly show a diagonal pattern, indicating poor generalization to other types of image distortions. From the bottom two rows, we see that both the combined model and our adversarially trained model are robust to distortions used to train the combined model (first five columns).

Figure~\ref{fig:noise_results} gives a more comprehensive comparison across a range of distortion levels. Our model reaches comparable performance on crop and dropout, outperforms the combined model on JPEG, and underperforms on Gaussian blur. For small distortion strengths, our accuracy is nearly identical to the combined model.  On all noise levels and distortion types, we outperform the identity model by a wide margin. 

In terms of visual quality, our model is less prone to small artifacts in flat regions of an image.  A qualitative comparison can be found in Figure~\ref{fig:samples}.

\subsection{Generalization to Unknown Distortions}
\label{subsec:unseen}
\begin{figure}[!ht]
\begin{center}
\vspace{-0 mm}
    \begin{tabular}{c c c c}
          \begin{subfigure}{.23\linewidth} \includegraphics[width=0.99\linewidth]{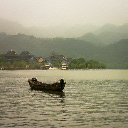} \end{subfigure}       
                 &  \hspace{-10pt} \begin{subfigure}{.23\linewidth} \includegraphics[width=0.99\linewidth]{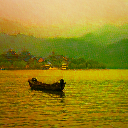} \end{subfigure}
                 & \hspace{-17pt} \begin{subfigure}{.23\linewidth} \includegraphics[width=0.99\linewidth]{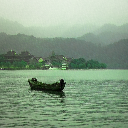} \end{subfigure}
                                  &  \hspace{-17pt} \begin{subfigure}{.16\linewidth} \includegraphics[width=0.99\linewidth]{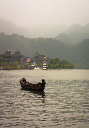} \end{subfigure}  \\
          Original & \hspace{-10pt} Saturation (5.0) & \hspace{-17pt} Hue(0.2) & \hspace{-17pt} Resize (0.7)  \\ 
          \begin{subfigure}{.23\linewidth} \includegraphics[width=0.99\linewidth]{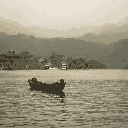} \end{subfigure}       
                 &  \hspace{-10pt} \begin{subfigure}{.23\linewidth} \includegraphics[width=0.99\linewidth]{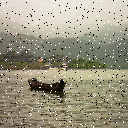} \end{subfigure}     
                 &  \hspace{-17pt}\begin{subfigure}{.23\linewidth} \includegraphics[width=0.99\linewidth]{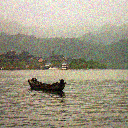} \end{subfigure} &\\
           GIF (16) & \hspace{-10pt} S\&P (0.05) & \hspace{-17pt} Gaussian (0.06) \\
    \end{tabular}
\end{center}
\caption{Visualization of additional image distortions.} \label{fig:unseen}
\end{figure}

A practical watermarking system must be robust to a wide range of image distortions, not just the distortions seen during training. Therefore, we compare the performance of our model and the combined model on a held-out set of commonly used image distortions. To attain better coverage, we choose six types of distortions, \textit{i.e.}, saturation, hue, resize, Gaussian noise, salt and pepper noise, GIF encoding from four broad categories: color adjustment, pixel-wise noise, geometric transformations, and compression. For each type of distortion, we evaluate the models on three different values of distortion strengths. Figure~\ref{fig:unseen} gives a visualization of the additional distortions.  We choose the range of distortion strength strong enough to differentiate the performance between different models, but in a regime where the distorted image still resembles the original.

\begin{table}[!ht]
    \begin{center}
    \begin{tabular}{c c c c c}
    \hline
         Method               & Identity     & Combined            & Ours    \\ \hline
         Gaussian Noise (0.06)       & 74.6         &  93.5               &    \textbf{95.6}          \\  
         Gaussian Noise (0.08)       & 67.7         &  87.2               &    \textbf{93.5}          \\      
         Gaussian Noise (0.10)       & 63.2         &  80.4               &    \textbf{89.5}          \\ \hline             
         Salt and Pepper (0.05)       & \textbf{99.1}            &  97.2               &    95.7          \\       
         Salt and Pepper (0.10)       & \textbf{93.1}            &  89.4               &    85.0          \\       
         Salt and Pepper (0.15)       & \textbf{83.4}            &  79.6               &    77.1          \\     \hline        
         Adjust Hue (0.2)            & 65.1         &  70.8             & \textbf{94.0}         \\
         Adjust Hue (0.4)            & 34.0         &  45.3             & \textbf{70.7}         \\  
         Adjust Hue (0.6)            & 18.1         &  28.8             & \textbf{42.4}         \\  \hline        
         Adjust Saturation (5.0)            & 96.3      &98.1             &\textbf{99.9}          \\          
         Adjust Saturation (10.0)            & 94.8      &96.0             &\textbf{99.6}          \\    
         Adjust Saturation (15.0)            & 93.4      &94.2             &\textbf{98.5}          \\     \hline   
         GIF (64)       & 87.1           &  96.5               &    \textbf{97.6}          \\       
         GIF (32)       & 76.8            &  93.4               &    \textbf{95.7}          \\           
         GIF (16)       & 65.0            &  88.6               &    \textbf{91.7}          \\          \hline            
         Resize Width (0.9)             & 99.3     &99.7             &\textbf{99.9}     \\        
         Resize Width (0.7)             & 85.3     &84.9             &\textbf{88.4}   \\         
         Resize Width (0.5)             & 66.5     &\textbf{67.3}             &67.1         \\  \hline   
         Average     &     78.37   &  84.26         &    \textbf{88.30}         \\ \hline
    \end{tabular}
    \end{center}
    \caption{Comparison of our model with HiDDeN identity model and combined noise model on additional image distortions. We report the bit accuracy of our model, the HiDDeN combined and identity model. When computing the average, results lower than $50\%$ are truncated to $50\%$ since they are no better than random chance.}
    \label{tab:bit-accuracy-comparison}
\end{table}

\begin{figure*}
\begin{center}
\vspace{-0 mm}
    \begin{tabular}{c c c c c c}
        Original &  \begin{subfigure}{.14\linewidth} \includegraphics[width=0.99\linewidth]{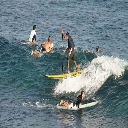} \end{subfigure}       
                 &  \begin{subfigure}{.14\linewidth} \includegraphics[width=0.99\linewidth]{./figures/samples/4_original.png} \end{subfigure}     
                 &  \begin{subfigure}{.14\linewidth} \includegraphics[width=0.99\linewidth]{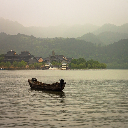} \end{subfigure}
                 &  \begin{subfigure}{.14\linewidth} \includegraphics[width=0.99\linewidth]{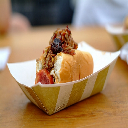} \end{subfigure}
                 &  \begin{subfigure}{.14\linewidth} \includegraphics[width=0.99\linewidth]{./figures/samples/47_original.png} \end{subfigure}          \vspace{10pt}       \\    
        HiDDeN   &  \begin{subfigure}{.14\linewidth} \includegraphics[width=0.99\linewidth]{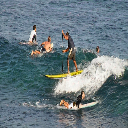} \end{subfigure}       
                 &  \begin{subfigure}{.14\linewidth} \includegraphics[width=0.99\linewidth]{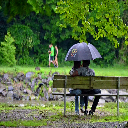} \end{subfigure}     
                 &  \begin{subfigure}{.14\linewidth} \includegraphics[width=0.99\linewidth]{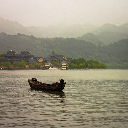} \end{subfigure}
                 &  \begin{subfigure}{.14\linewidth} \includegraphics[width=0.99\linewidth]{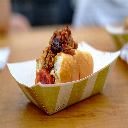} \end{subfigure}
                 &  \begin{subfigure}{.14\linewidth} \includegraphics[width=0.99\linewidth]{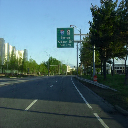} \end{subfigure}          \vspace{10pt}       \\   
        Ours     &  \begin{subfigure}{.14\linewidth} \includegraphics[width=0.99\linewidth]{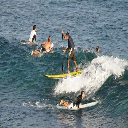} \end{subfigure}       
                 &  \begin{subfigure}{.14\linewidth} \includegraphics[width=0.99\linewidth]{./figures/samples/4_encoded_adv.png} \end{subfigure}     
                 &  \begin{subfigure}{.14\linewidth} \includegraphics[width=0.99\linewidth]{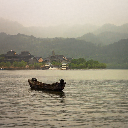} \end{subfigure}
                 &  \begin{subfigure}{.14\linewidth} \includegraphics[width=0.99\linewidth]{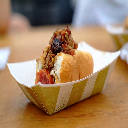} \end{subfigure}
                 &  \begin{subfigure}{.14\linewidth} \includegraphics[width=0.99\linewidth]{./figures/samples/47_encoded_adv.png} \end{subfigure}            \vspace{10pt}     \\                   
        Diff (HiDDeN)   &  \begin{subfigure}{.14\linewidth} \includegraphics[width=0.99\linewidth]{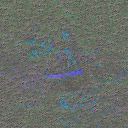} \end{subfigure}       
                 &  \begin{subfigure}{.14\linewidth} \includegraphics[width=0.99\linewidth]{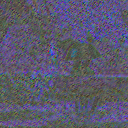} \end{subfigure}     
                 &  \begin{subfigure}{.14\linewidth} \includegraphics[width=0.99\linewidth]{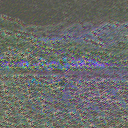} \end{subfigure}
                 &  \begin{subfigure}{.14\linewidth} \includegraphics[width=0.99\linewidth]{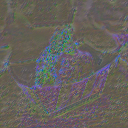} \end{subfigure}
                 &  \begin{subfigure}{.14\linewidth} \includegraphics[width=0.99\linewidth]{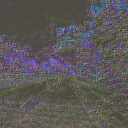} \end{subfigure}          \vspace{10pt}       \\                  
        Diff (Ours)     &  \begin{subfigure}{.14\linewidth} \includegraphics[width=0.99\linewidth]{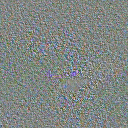} \end{subfigure}       
                 &  \begin{subfigure}{.14\linewidth} \includegraphics[width=0.99\linewidth]{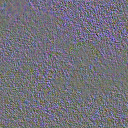} \end{subfigure}     
                 &  \begin{subfigure}{.14\linewidth} \includegraphics[width=0.99\linewidth]{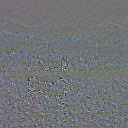} \end{subfigure}
                 &  \begin{subfigure}{.14\linewidth} \includegraphics[width=0.99\linewidth]{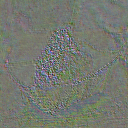} \end{subfigure}
                 &  \begin{subfigure}{.14\linewidth} \includegraphics[width=0.99\linewidth]{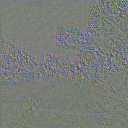} \end{subfigure}            \vspace{10pt}     \\                    
    \end{tabular}
\end{center}
\caption{Samples of encoded and cover images for the watermarking algorithm. \textbf{First row:} Cover image with no embedded message. \textbf{Second row:} Encoded image from HiDDeN combined distortion model. \textbf{Third row:} Encoded images from our model. \textbf{Fourth row:} Normalized difference of the encoded image and cover image for the HiDDeN combined model. \textbf{Fifth row:} Normalized difference for our model. } \label{fig:samples}
\end{figure*}

Table \ref{tab:bit-accuracy-comparison} reports the bit accuracy of our model on these additional distortions. Overall, our model performs better on the unknown distortions, especially on the category of color change. We also note that the overall variance of bit accuracy across distortions is less compared to both the identity and combined model, indicating a more stable performance across different types of distortions. Furthermore, we see that the performance gap of the combined model and the identity model shrinks on these unknown distortions, which aligns with the intuition that generalization issue persists even when training with a combination of distortions.

\subsection{Detailed Analysis}
\label{subsec:detailed-analysis}
\subsubsection{Ablation Study}
Table~\ref{tab:ablation} reports the individual effect of channel coding and adversarial training. We see that adversarial training contributes to a large portion of the model robustness, while channel coding further boosts performance in terms of accuracy. Table~\ref{tab:ablation} also shows that channel coding alone does not provide enough robustness without a robust watermarking model. However, combined with adversarial training channel coding further boosts the performance of the watermarking system especially if the bit accuracy is already high.

\begin{table}[!ht]
    \begin{center}
    \begin{tabular}{p{0.15\linewidth}p{0.15\linewidth}p{0.15\linewidth}p{0.15\linewidth}p{0.12\linewidth}}
    \hline
                       & JPEG (Q=50)     & Crop (p=0.09)               &  Blur ($\sigma$=1.0)   & Dropout (p=0.3)\\ \hline
         Identity         & 50.2     & 53.0                 &59.6    &  81.3      \\                       
         Channel          & 51.3     & 60.5                 &50.2       & 90.3 \\
         Adv.      & \textbf{85.0}     & 90.6                 &86.2    &  95.0   \\
         Both             & 81.7    & \textbf{93.5}                 &\textbf{92.8}    &  \textbf{97.9}  \\ \hline     
    \end{tabular}
    \end{center}
    \caption{Model ablation study. We report the bit accuracy for models trained with only channel coding, only adversarial training, both, and the identity model. For models trained with only adversarial training, the input message length to the watermarking model is $30$ instead of $120$.}
    \label{tab:ablation}
 
\end{table}


\subsubsection{Attack Complexity}
We study the effect of varying the complexity and architecture of the attack network $G_{adv}$. On top of adjusting the network size and depth, we also consider two variants of the attack network: the \emph{residual} network (Res) where we add a skip connection from the input, and a \emph{capped} network (Capped) where we limit the maximum pixel difference by setting $G_{adv}(I) = I + \epsilon \hspace{3pt}\text{tanh}(\text{CNN}(I))$. We also report the results from the fast gradient sign method (FGSM) for completeness.

\begin{table}[!ht]
    \begin{center}
    \begin{tabular}{p{0.23\linewidth}p{0.13\linewidth}p{0.13\linewidth}p{0.13\linewidth}p{0.10\linewidth}}
    \hline
                       & JPEG (Q=50)     & Dropout (p=0.3)               & Blur ($\sigma$=1.0)   & Acc. (adv.)\\ \hline
         Conv  \footnotesize{(3,16)}         &  81.7     & 97.9            & \textbf{92.8}   & 90.6 \\  
         Conv \footnotesize{(3,32)}         &  80.5     & \textbf{98.0}    & 84.9   & 78.7  \\ 
         Conv \footnotesize{(3,32,32)}      &  75.0         & 95.3         &81.5      & 72.0  \\ 
         Res \footnotesize{(3,16)}           & \textbf{84.5} & 96.3        &86.3 & 95.0       \\ 
         Capped \footnotesize{(0.03)}       &  57.3        & 93.9          &77.3  & 96.5       \\           
         Capped \footnotesize{(0.06)}      &  53.2        & 94.6          &78.1  & \textbf{99.5}       \\           
         FGSM                & 50.1     & 86.2       & 50.1    & 98.0    \\   \hline  
    \end{tabular}
    \end{center}
    \caption{Performance when varying attack network complexity. Each row corresponds to models trained with a different configuration of attack network. The first three columns show the bit accuracy on various image distortions. The last column shows the bit accuracy on adversarial message $X'_{adv}$. }
    \label{tab:attack-network-tradeoff}
\end{table}
From Table~\ref{tab:attack-network-tradeoff}, we observe that the bit accuracy on the adversarial example decreases as the attack network complexity increases, causing a slight degradation in the final result. Capping the attack network yielded poor results on JPEG and Gaussian blur, indicating that this approach over-restricts the attack network. The residual network yielded very similar performance to the regular convolutional model, slightly underperforming on Gaussian blur. Finally, FGSM yielded poor results on all of the distortions, since the image networks quickly overfits to this specific type of distortion.


\section{Conclusion}
\label{sec:conclusion}
We propose a distortion agnostic watermarking method that does not explicitly model the image distortion at training time. Our method consists of two core components, adversarial training and channel coding, to improve the robustness of our system. Compared with conventional methods of improving model robustness, our methods do not require the explicit modeling of the image distortions at training time. Through empirical evaluations, we validate that our model reaches comparable performance to the combined distortion model on distortions seen during training, and better generalization to unseen distortions. In future work, we would like to improve upon our current methodology to further increase model robustness, and explore deeper the connections between watermarking and adversarial attacks.

{\small
\bibliographystyle{ieee_fullname}
\bibliography{reference}
}

\onecolumn
\section{Appendix}

\subsection{Training Details}

We list the hyper-parameters used for training the watermarking model. For our model, we set $\alpha_1^I =18.0, \alpha_2^I=0.01, \alpha^M=0.3, \alpha_1^{adv}=15.0, \alpha_2^{adv}=1.0, \alpha_W^{adv}=0.2$, and $num\_iter=5$. For the HiDDeN combined model and identity model, we set $\alpha_1^I =6.0, \alpha_2^I=0.01, \alpha^M=1.0$. The message size for our watermarking model is $120$ instead of $30$, due to the addition of the channel coding layer.  We use the same network architecture as in HiDDeN. Namely, the input image $I_{co}$ is first processed by 4 $3\times3$ Conv-BN-ReLU blocks with 64 units per layer. This is then concatenated along the channel dimension with an $H \times W$ spatial repetition of the input message. The combined blocks are then passed to two additional Conv-BN-ReLU blocks to produce the encoded image. For the encoder, we symmetrically pad the input image and use 'VALID' padding for all convolution operations to reduce boundary artifacts of the encoded image. The encoded image is clipped to $[0, 1]$ before passing to the decoder. The decoder consists of seven $3\times3$ Conv-BN-ReLU layers of size, where the last two layers have stride 2. A global pooling operation followed by a fully-connected layer is used to produce the decoded message.

For both our model and the combined model, the training warm-starts from a pre-trained HiDDeN identity model and stops at $250$k iterations. We use ADAM with a learning rate of $1e-3$ for all models. 

For the channel model, we use a two fully connected layers with 512 units each, and train with BSC noise where the noise strength is uniformly sample from $[0, 0.3]$.

\subsection{Encoded Image Samples}

\begin{figure*}
\begin{center}
\vspace{-0 mm}
    \begin{tabular}{c c c c c c}
    Original & Ours & Diff (Ours) & Combined & Diff (Combined) \\
         \begin{subfigure}{.19\linewidth} \includegraphics[width=0.99\linewidth]{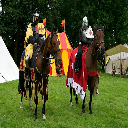} \end{subfigure}       
                 &  \begin{subfigure}{.19\linewidth} \includegraphics[width=0.99\linewidth]{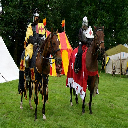} \end{subfigure}     
                 &  \begin{subfigure}{.19\linewidth} \includegraphics[width=0.99\linewidth]{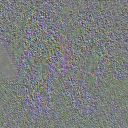} \end{subfigure}
                 &  \begin{subfigure}{.19\linewidth} \includegraphics[width=0.99\linewidth]{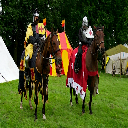} \end{subfigure}
                 &  \begin{subfigure}{.19\linewidth} \includegraphics[width=0.99\linewidth]{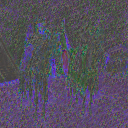} \end{subfigure}    \vspace{10pt}    \\    
         \begin{subfigure}{.19\linewidth} \includegraphics[width=0.99\linewidth]{./supplement/253_original.png} \end{subfigure}       
                 &  \begin{subfigure}{.19\linewidth} \includegraphics[width=0.99\linewidth]{./supplement/253_encoded_adv.png} \end{subfigure}     
                 &  \begin{subfigure}{.19\linewidth} \includegraphics[width=0.99\linewidth]{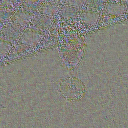} \end{subfigure}
                 &  \begin{subfigure}{.19\linewidth} \includegraphics[width=0.99\linewidth]{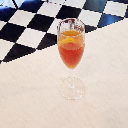} \end{subfigure}
                 &  \begin{subfigure}{.19\linewidth} \includegraphics[width=0.99\linewidth]{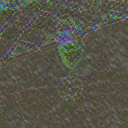} \end{subfigure}    \vspace{10pt}    \\    
         \begin{subfigure}{.19\linewidth} \includegraphics[width=0.99\linewidth]{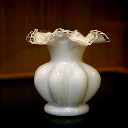} \end{subfigure}       
                 &  \begin{subfigure}{.19\linewidth} \includegraphics[width=0.99\linewidth]{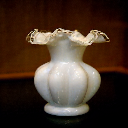} \end{subfigure}     
                 &  \begin{subfigure}{.19\linewidth} \includegraphics[width=0.99\linewidth]{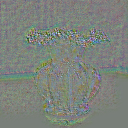} \end{subfigure}
                 &  \begin{subfigure}{.19\linewidth} \includegraphics[width=0.99\linewidth]{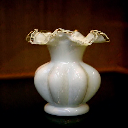} \end{subfigure}
                 &  \begin{subfigure}{.19\linewidth} \includegraphics[width=0.99\linewidth]{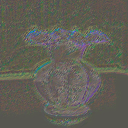} \end{subfigure}     \vspace{10pt}       \\       
         \begin{subfigure}{.19\linewidth} \includegraphics[width=0.99\linewidth]{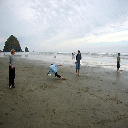} \end{subfigure}       
                 &  \begin{subfigure}{.19\linewidth} \includegraphics[width=0.99\linewidth]{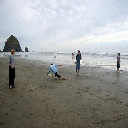} \end{subfigure}     
                 &  \begin{subfigure}{.19\linewidth} \includegraphics[width=0.99\linewidth]{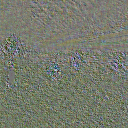} \end{subfigure}
                 &  \begin{subfigure}{.19\linewidth} \includegraphics[width=0.99\linewidth]{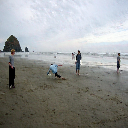} \end{subfigure}
                 &  \begin{subfigure}{.19\linewidth} \includegraphics[width=0.99\linewidth]{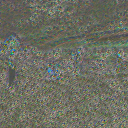} \end{subfigure}     \vspace{10pt}       \\  
         \begin{subfigure}{.19\linewidth} \includegraphics[width=0.99\linewidth]{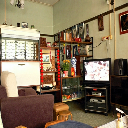} \end{subfigure}       
                 &  \begin{subfigure}{.19\linewidth} \includegraphics[width=0.99\linewidth]{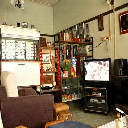} \end{subfigure}     
                 &  \begin{subfigure}{.19\linewidth} \includegraphics[width=0.99\linewidth]{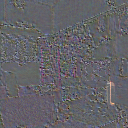} \end{subfigure}
                 &  \begin{subfigure}{.19\linewidth} \includegraphics[width=0.99\linewidth]{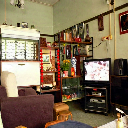} \end{subfigure}
                 &  \begin{subfigure}{.19\linewidth} \includegraphics[width=0.99\linewidth]{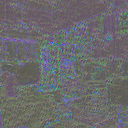} \end{subfigure}     \vspace{10pt}       \\  
    \end{tabular}
\end{center}
\caption{Samples of encoded image from HiDDeN and our model. } \label{fig:samples}
\end{figure*}

\begin{figure*}
\begin{center}
\vspace{-0 mm}
    \begin{tabular}{c c c c c c}
    Original & Ours & Diff (Ours) & Combined & Diff (Combined) \\
         \begin{subfigure}{.19\linewidth} \includegraphics[width=0.99\linewidth]{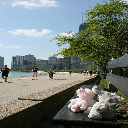} \end{subfigure}       
                 &  \begin{subfigure}{.19\linewidth} \includegraphics[width=0.99\linewidth]{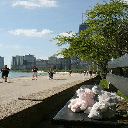} \end{subfigure}     
                 &  \begin{subfigure}{.19\linewidth} \includegraphics[width=0.99\linewidth]{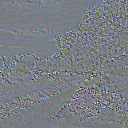} \end{subfigure}
                 &  \begin{subfigure}{.19\linewidth} \includegraphics[width=0.99\linewidth]{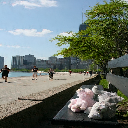} \end{subfigure}
                 &  \begin{subfigure}{.19\linewidth} \includegraphics[width=0.99\linewidth]{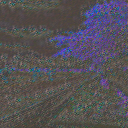} \end{subfigure}     \vspace{10pt}       \\   
         \begin{subfigure}{.19\linewidth} \includegraphics[width=0.99\linewidth]{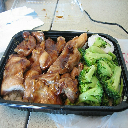} \end{subfigure}       
                 &  \begin{subfigure}{.19\linewidth} \includegraphics[width=0.99\linewidth]{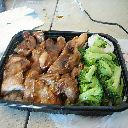} \end{subfigure}     
                 &  \begin{subfigure}{.19\linewidth} \includegraphics[width=0.99\linewidth]{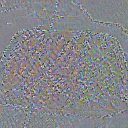} \end{subfigure}
                 &  \begin{subfigure}{.19\linewidth} \includegraphics[width=0.99\linewidth]{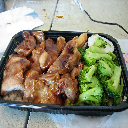} \end{subfigure}
                 &  \begin{subfigure}{.19\linewidth} \includegraphics[width=0.99\linewidth]{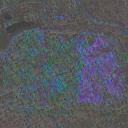} \end{subfigure}     \vspace{10pt}       \\  
         \begin{subfigure}{.19\linewidth} \includegraphics[width=0.99\linewidth]{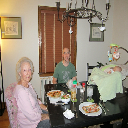} \end{subfigure}       
                 &  \begin{subfigure}{.19\linewidth} \includegraphics[width=0.99\linewidth]{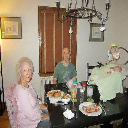} \end{subfigure}     
                 &  \begin{subfigure}{.19\linewidth} \includegraphics[width=0.99\linewidth]{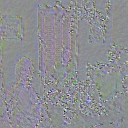} \end{subfigure}
                 &  \begin{subfigure}{.19\linewidth} \includegraphics[width=0.99\linewidth]{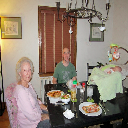} \end{subfigure}
                 &  \begin{subfigure}{.19\linewidth} \includegraphics[width=0.99\linewidth]{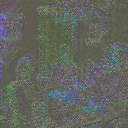} \end{subfigure}     \vspace{10pt}       \\   
         \begin{subfigure}{.19\linewidth} \includegraphics[width=0.99\linewidth]{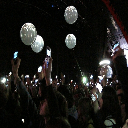} \end{subfigure}       
                 &  \begin{subfigure}{.19\linewidth} \includegraphics[width=0.99\linewidth]{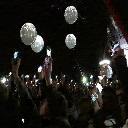} \end{subfigure}     
                 &  \begin{subfigure}{.19\linewidth} \includegraphics[width=0.99\linewidth]{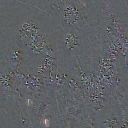} \end{subfigure}
                 &  \begin{subfigure}{.19\linewidth} \includegraphics[width=0.99\linewidth]{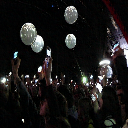} \end{subfigure}
                 &  \begin{subfigure}{.19\linewidth} \includegraphics[width=0.99\linewidth]{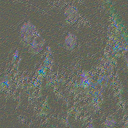} \end{subfigure}     \vspace{10pt}       \\    
         \begin{subfigure}{.19\linewidth} \includegraphics[width=0.99\linewidth]{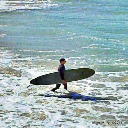} \end{subfigure}       
                 &  \begin{subfigure}{.19\linewidth} \includegraphics[width=0.99\linewidth]{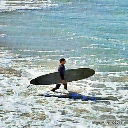} \end{subfigure}     
                 &  \begin{subfigure}{.19\linewidth} \includegraphics[width=0.99\linewidth]{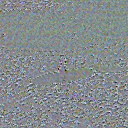} \end{subfigure}
                 &  \begin{subfigure}{.19\linewidth} \includegraphics[width=0.99\linewidth]{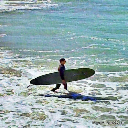} \end{subfigure}
                 &  \begin{subfigure}{.19\linewidth} \includegraphics[width=0.99\linewidth]{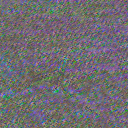} \end{subfigure}     \vspace{10pt}       \\                  
    \end{tabular}
\end{center}
\caption{More samples of encoded image from HiDDeN and our model. } \label{fig:samples}
\end{figure*}
\newpage

\subsection{Adversarial Example Samples}

\begin{figure*}[!ht]
\begin{center}
\vspace{-10pt}
    \begin{tabular}{c c c}
    Encoded Image & Adversarial Example & Difference \\
         \begin{subfigure}{.22\linewidth} \includegraphics[width=0.99\linewidth]{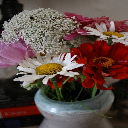} \end{subfigure}       
                 &  \begin{subfigure}{.22\linewidth} \includegraphics[width=0.99\linewidth]{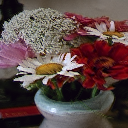} \end{subfigure}     
                 &  \begin{subfigure}{.22\linewidth} \includegraphics[width=0.99\linewidth]{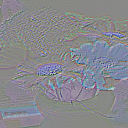} \end{subfigure}  \vspace{10pt}       \\   
         \begin{subfigure}{.22\linewidth} \includegraphics[width=0.99\linewidth]{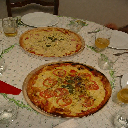} \end{subfigure}       
                 &  \begin{subfigure}{.22\linewidth} \includegraphics[width=0.99\linewidth]{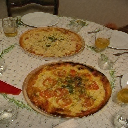} \end{subfigure}    
                 &  \begin{subfigure}{.22\linewidth} \includegraphics[width=0.99\linewidth]{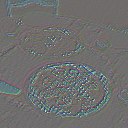} \end{subfigure}  \vspace{10pt}       \\  
         \begin{subfigure}{.22\linewidth} \includegraphics[width=0.99\linewidth]{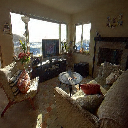} \end{subfigure}       
                 &  \begin{subfigure}{.22\linewidth} \includegraphics[width=0.99\linewidth]{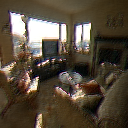} \end{subfigure}    
                 &  \begin{subfigure}{.22\linewidth} \includegraphics[width=0.99\linewidth]{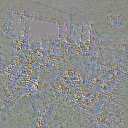} \end{subfigure}  \vspace{10pt}       \\        
         \begin{subfigure}{.22\linewidth} \includegraphics[width=0.99\linewidth]{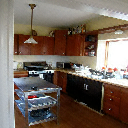} \end{subfigure}       
                 &  \begin{subfigure}{.22\linewidth} \includegraphics[width=0.99\linewidth]{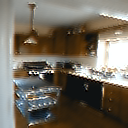} \end{subfigure}    
                 &  \begin{subfigure}{.22\linewidth} \includegraphics[width=0.99\linewidth]{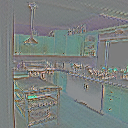} \end{subfigure}  \vspace{10pt}       \\ 
         \begin{subfigure}{.22\linewidth} \includegraphics[width=0.99\linewidth]{./supplement/adv_examples/encoded_image_5.png} \end{subfigure}       
                 &  \begin{subfigure}{.22\linewidth} \includegraphics[width=0.99\linewidth]{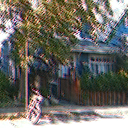} \end{subfigure}    
                 &  \begin{subfigure}{.22\linewidth} \includegraphics[width=0.99\linewidth]{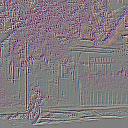} \end{subfigure}         \\                  
    \end{tabular}
\end{center}
\caption{Samples of adversarial examples generated by the attack network. } \label{fig:adv-samples}
\end{figure*}

\end{document}